\documentclass[12pt]{article}
\usepackage{latexsym}
\usepackage{amssymb}
\usepackage[cp1251]{inputenc}

\title{One-loop effective action in ${\cal N}=2$ supersymmetric massive
Yang-Mills theory}

\author{ I.L.
Buchbinder$^1$\footnote{joseph@tspu.edu.ru},  N.G.
Pletnev$^2$\footnote{pletnev@math.nsc.ru}}

\date{\it
$^1$ Department of Theoretical Physics\\
Tomsk State Pedagogical University\\ Tomsk 634041, Russia\\
$^2$Department of
Theoretical Physics\\ Institute of Mathematics, Novosibirsk , \\
630090, Russia\\}

\begin{document}

\maketitle

\begin{center}
{\bf Abstract}
\end{center}
We consider the ${\cal N}=2$ supersymmetric theory of the massive
Yang-Mills field formulated in the ${\cal N}=2$ harmonic superspace.
The various gauge-invariant forms of writing the mass term in the
action (in particular, using the Stueckelberg superfield), which
result in dual formulations of the theory, are presented. We develop
a gauge-invariant and explicitly supersymmetric scheme of the loop
off-shell expansion of the superfield effective action. In the
framework of this scheme, we calculate gauge-invariant and
explicitly ${\cal N}=2$ supersymmetric one-loop counterterms
including new counterterms depending on the Stueckelberg superfield.
Component structure of one of these counterterms is analyzed.

\section{Introduction}
Study of quantum aspects of massive non-Abelian Yang-Mills
theories has long history (see, e.g., \cite{histor}). It is very
well known that to understand particle phenomenology, the massive
degrees of freedom of vector bosons must be taken into account.
However the mass terms in vector field Lagrangian violates a gauge
invariance of massless theory.

Several different mechanisms for generating vector field mass are
currently known that are compatible with the gauge invariance.
Their common feature it is increasing the number of physical
degrees of freedom in comparison with the massless theory. Of
course, the main commonly accepted paradigm of the standard model
is the mechanism of spontaneous symmetry breaking in which
additional physical degrees of freedom are due to scalar Higgs
fields. However for phenomenological aims is useful to consider
other mechanisms for generating the boson and fermion masses in
gauge theories such that no additional physical fields appear in
the Lagrangian.

The most popular alternative to the Higgs mechanism is the model
based on the massive Yang-Mills theory. In such a model, the gauge
invariance is attained by introducing the real pseudoscalar
auxiliary Stueckelberg field \cite{stueck}, which corresponds to
coupling the Yang-Mills field to a gauge nonlinear sigma model. In
the unitary gauge, this field is absorbed by the longitudinal
component of the massive vector field (see \cite{ruegg} for a
comprehensive review and reference list; we mention \cite{recent}
among the numerous latest publications).

In addition to models with the Stueckelberg fields, non-Abelian
vector-tensor gauge theories with topological constraints can be
considered \cite{townsend}. All such theories are classically
equivalent to non-Abelian theories of massive vector fields with
the Stueckelberg fields. The same degrees of freedom can be
described by either of the two dual representations. Depending on
a problem context, one of the formulations can be more convenient,
and both formulations and their interrelations are therefore worth
studying.\footnote{The quantum equivalence of such different dual
formulations is a more delicate problem requiring a separate
investigation for each concrete case (see, e.g., \cite{quant}).}
We mention that such models appear naturally in the low-energy
limit of the superstring theory and also in the context of
supergravity in higher dimensions. For example, degrees of freedom
of the massive skew-symmetric tensor field related to the
mechanism of natural spontaneous supersymmetry breaking appear
naturally in the recently found compactifications of the type-II
superstring on the Calabi-Yau manifolds in the presence of
nontrivial flows of the 3-form (see, e.g., \cite{grana}). This
revived the interest in a more detailed study of massive ${\cal
N}=1$ and ${\cal N}=2$ tensor multiplets and their relations to
scalar and vector multiplets. We note that such the relations play
an important role in the mechanism of anomaly cancellation in
superstring models.

The construction of the ${\cal N}=1$ supersymmetric massive tensor
multiplet as the version dual to the massive vector multiplet has
long been known (see, e.g., \cite{idea}). In the ${\cal N}=2$
supersymmetry, the strength of the skew-symmetric tensor fields is
contained in the tensor multiplet $G^{++}$ defined on the analytic
subspace of the harmonic ${\cal N}=2$ superspace restricted by
constraints \cite{gikos}, \cite{gios}. The action contains only
$G^{++}$ in the case of massless tensor multiplet, but if the
skew-symmetric tensor acquires mass, then the gauge invariance
results in the relation between Stueckelberg fields and the vector
multiplet \cite{kuz_mass}. Studying the quantum properties of dual
realizations of the same supersymmetry representation is
especially interesting.

Here, we consider the quantum properties of the ${\cal N}=2$
massive Yang-Mills field theory with the Stueckelberg fields. This
model is a direct ${\cal N}=2$ supersymmetrization of the ${\cal
N}=0$ nonsupersymmetric massive Yang-Mills theory in the
Kunimasa-Goto formalism \cite{kg}. Several aspects of this problem
were already considered in \cite{kh}, where it was found that the
theory is finite in the second order in the dimensionless
Yang-Mills coupling constant $g^2$ and that the massive term is
not renormalized, but the theory then becomes nonrenormalizable in
the sector containing the dimension full coupling constant
$\frac{m^2}{g^2}$. It was concluded from this that the theory is
finite in all orders of the loop expansion in the vector multiplet
sector.

But we note that already on the classical level, the action of the
${\cal N}=2$ massive Yang{Mills theory has the form of an infinite
series containing all orders of the vector multiplet potential
$V^{++}$ in the framework of the harmonic superspace formalism
\cite{gikos}. Moreover, the sigma-model Lagrangian of the
Stueckelberg superfield is itself highly nonlinear. To analyze the
quantum properties of the theory in a gauge invariant way even on
the one-loop level, we therefore cannot restrict ourself to
considering only the simplest diagrams resulting in
gauge-noninvariant counterterms and must instead sum over all
one-loop diagrams with all possible external legs for the
effective action. In the framework of the standard noncovariant
diagram technique, this problem seems to be very difficult
technically, if not impossible.

Here, to construct the effective action, we use the formulation of
the ${\cal N}=2$ supersymmetric Yang-Mills field theory and the
corresponding Stueckelberg formalism in the harmonic superspace
\cite{kh} and the background field method \cite{backgr}, which
allows effectively summing all the diagrams with the increasing
number of insertions of the external lines. Our conclusions are
ideologically close to the results in \cite{shiza}, \cite{yukaf},
where the problem of constructing off-mass-shell invariant
counterterms for the nonsupersymmetric (${\cal N}=0$) massive
Yang-Mills theory was solved. To preserve the gauge invariance at
all calculation stages, we use the invariant perturbation theory
developed in models of principal chiral fields long ago \cite{volk}.

The paper is organized as follows. In Sec. 2, we describe the
formulation of the ${\cal N}=2$ supersymmetric theory of the massive
Yang-Mills field in the harmonic superspace taking into account the
Stueckelberg superfield. Excluding nonphysical degrees of freedom,
we obtain an explicitly gauge-invariant nonlocal expression for the
mass term in the Lagrangian. It is expected that the dual relation
to the theory of ${\cal N}=2$ massive tensor multiplet
\cite{kuz_mass} becomes more transparent just in this formulation.
In Sec. 3, we discuss the procedure for constructing the effective
action based on the ${\cal N}=2$ supersymmetric background field
method and indicate the special features of using this method in the
theory under consideration. Sec. 4 is devoted to calculating the
one-loop divergences in the effective action. There, we first
present gauge-invariant and explicitly ${\cal N}=2$ supersymmetric
counterterms depending on the Stueckelberg superfield. In Sec. 5, we
discuss the derivation of the component structure of the bosonic
sector of one of these counterterms.

\section{The ${\cal N}=2$ supersymmetric theory of the massive Yang-Mills
field in the harmonic superspace}

The formulation of the ${\cal N}=2$ supersymmetric field theories in
terms of unrestricted superfields defined on an analytic subspace of
the harmonic superspace \cite{gikos}, \cite{gios} turns out to be
exceptionally useful for investigating the quantum effects (see,
e.g., \cite{backgr}, \cite{effect}). The concept of the harmonic
${\cal N}=2$ superfield was introduced in \cite{gikos}; it consists
in enhancing the standard ${\cal N}=2$ superspace with the
coordinates $z^M=(x^m, \theta^\alpha_i, \bar\theta^i_{\dot\alpha})
(i=1,2)$, by adding the spherical harmonics $u^{\pm}_i$
parameterizing the two-dimensional sphere $S^2=SU(2)/U(1)$:
$u^{+i}u^-_i=1, \overline{u^{+i}}=u^-_i.$ The main advantage of
using the harmonic superspace is that unconstrained superfields of
matter hypermultiplets and those of the vector Yang-Mills field
multiplet are defined in the analytic subspace with the coordinates
$\zeta^M=(x^m_A, \theta^{+\alpha}, \bar\theta^+_{\dot\alpha},
u^{\pm}_i)$ in which the so-called analytic basis is closed under
the transformations of the ${\cal N}=2$ supersymmetry:
$$
x^m_A=x^m-i\theta^+\sigma^m\bar\theta^-
-i\theta^-\sigma^m\bar\theta^+, \quad
\theta^{\pm}_\alpha=u^{\pm}_i\theta^i_\alpha, \quad
\bar\theta^{\pm}_{\dot\alpha}=u^{\pm}_i\bar\theta^i_{\dot\alpha}.
$$
The ${\cal N}=2$ vector multiplet with a finite number of physical
and auxiliary component fields but with an infinite number of gauge
degrees of freedom is described by a real analytic superfield
$V^{++}=V^{++}_a T_a$ taking values in the Lie algebra of the gauge
group. The hypermultiplets $\omega$ and $q^+$ containing off-shell
an infinite number of auxiliary fields and transforming on a
representation R of the gauge group are determined by the analytic
superfields $\omega(\zeta)$, $q^+(\zeta)$, and their conjugate
$\tilde{q}^+(\zeta)$ (see \cite{gios} for the definition of the
generalized conjugation as a composition of the standard conjugation
and the antipodal mapping on the two-sphere). The scalar component
fields $\omega(x_A)$ and $\omega^{(ij)}(x_A)$ of the
$\omega$-multiplet, which are the respective isoscalar and
isotriplet of the $SU(2)$ group of internal isomorphisms of the
supersymmetry algebra, and the doublet of the Weyl fermions
$\psi^i_\alpha, \bar\psi_{i\, \dot\alpha}$ appear as lower
components in the expansion of $\omega(\zeta)$ in powers of
$\theta^+, \bar\theta^+$ and $u^{\pm}_i$ . Other ${\cal N}=2$ matter
multiplets with a finite number of auxiliary fields are described by
analytic superfields subject to proper harmonic constraints. The
vector ${\cal N}=2$ potential $V^{++}$ satisfies the reality
constraint with respect to the generalized conjugation
$\widetilde{V^{++}}=V^{++}$ and transforms as $\delta V^{++}=-{\cal
D}^{++}\lambda$ under the gauge transformations, where $\lambda$ is
an arbitrary real analytical superfield and ${\cal D}^{++}$ is the
covariant harmonic derivative in the analytic basis $${\cal
D}^{++}=D^{++}+iV^{++}=e^{ib(z,u)}D^{++}e^{-ib(z,u)},$$
$$D^{++}=u^{+i}\frac{\partial}{\partial
u^{-i}}-2i\theta^+\sigma^m\bar\theta^+\frac{\partial}{\partial
x^m_A}+\theta^{+\alpha}\frac{\partial}{\partial
\theta^{-\alpha}}+\bar\theta^{+\dot\alpha}\frac{\partial}{\partial
\bar\theta^{-\dot\alpha}}$$ and $b(z,u)$ is the so-called gauge
bridge. This gauge freedom allows eliminating an infinite number
of auxiliary fields by choosing the Wess-Zumino gauge in which the
analytic superfield $V^{++}$ contains a finite number of physical
and auxiliary fields. As shown in \cite{gikos}, \cite{gios}, all
the geometric characteristics, such as the field strength, can be
expressed in terms of a unique unrestricted potential
$V^{++}(\zeta,u)$.

We are not going to discuss the ${\cal N}=2$ supersymmetric
Yang-Mills field theory (see \cite{gios}) in detail and only present
the action of the non-Abelian vector multiplet. On-shell this
multiplet consists of the following component fields: the vector
field $A_m(x)$, the complex scalar field $M(x) +iN(x)$, the Majorana
isodoublet of spinors $\lambda^i_\alpha(x)$ and $
\bar\lambda^i_{\dot\alpha}(x)$ and the triplet of auxiliary fields
$F^{ij}(x)$. Off-shell, this multiplet is given by the superfield
potential $V^{++}$ taking values in the Lie algebra of the gauge
group. The corresponding action is:
\begin{equation}\label{class}
S
 =\frac{1}{2g^2}\mbox{tr}\sum_{n=2}^\infty
\frac{(-i)^n}{n}\int d^{12}z du_{1} ... du_n\frac{V^{++}(z,u_1)...
V^{++}(z,u_n)}{(u^+_1u^+_2)...(u^+_nu^+_1)}=-\frac{1}{2g^2}\mbox{tr}\int
d^8zW^2,
\end{equation}
where we use the harmonic distributions $\frac{1}{u^+_1u^+_2}$ or,
in other words, the Green's functions on the sphere
$G^{(-1,-1)}(u_1,u_2)$, which satisfy the equation
$\partial^{++}G^{(-1,-1)}(u_1,u_2)=\delta^{(1,-1)}(u_1,u_2).$ The
rules of differentiation with respect to harmonics and integration
over harmonics were defined in the pioneering papers \cite{gikos},
\cite{gios}.

The harmonic-independent chiral superfield strength
$W=-\frac14(\bar{D}^+)^2V^{--}$  is determined in terms of the
nonanalytic superfield $$V^{--}(z,u)=\int
du'\frac{e^{ib(z,u)}e^{-ib(z,u')}V^{++}(z,u')}{(u^+u^{' +})^2},$$
satisfying the zero-curvature
equation:\begin{equation}\label{zerocurve}
D^{++}V^{--}-D^{--}V^{++}+i[V^{++},V^{--}]=0.
\end{equation}
In the $\lambda$-basis, action (\ref{class}) is invariant under
the gauge transformations
\begin{equation}\label{gaugetr}
{}^g V^{++}=e^{i\lambda}(V^{++}-iD^{++})e^{-i\lambda}, \quad
V^{++}=V^{++}_a T_a,
\end{equation}
$$
\lambda=\lambda(\zeta,u)=\bar\lambda(\zeta,u), \quad
\lambda=\lambda_a T_a.
$$
where $T_a$ are the generators of the gauge group given by the
formulas:
$$
[T_a,T_b]=if_{abc}T_c, \quad \mbox{tr}(T_aT_b)=\delta_{ab},
$$
and the superfield gauge parameter $\lambda(\zeta,u)$ is a real
analytic superfield.

We consider a construction of  the gauge-invariant expression for
the mass term in the superfield action. For this, we use the known
Kunimasa-Goto formalism \cite{kg}, \cite{ruegg}, developed for
describing gauge field masses in the ${\cal N}=0$ Yang-Mills theory.
In the ${\cal N}=2$ superfield description, this formalism requires
introducing the additional Goldstone $\omega$-hypermultiplet in the
adjoint representation of the gauge group. The corresponding mass
term in the action is given as follows:
\begin{equation}\label{mass1}
S_m=-\frac{m^2}{2g^2}\mbox{tr}\int d\zeta^{(-4)}du
\{\Omega^{-1}(V^{++}-iD^{++})\Omega\}^2,
\end{equation}
where $\Omega= \Omega(\omega)= e^{-i\omega}$. In such a form of
writing,\footnote{Another useful form of writing (\ref{mass1}) is
$$S_m=-\frac{m^2}{2g^2}\mbox{tr}\int d\zeta^{(-4)}du
(V^{++}-i\Omega^{-1}(D^{++}\Omega+i[V^{++}, \Omega]))^2,$$ in
which the specific structure of the gauge sigma model is more
transparent.} the mass term is explicitly invariant under the
simultaneous transformations (\ref{gaugetr}) and the
transformations
\begin{equation}\label{transomeg}
{}^g\Omega=e^{i\lambda}\Omega,
\end{equation}
where ${}^g\Omega = g\Omega$ and $g$ is the gauge group element.
We note that mass term (\ref{mass1}) is also invariant under the
global right transformations $g_R$: $(V^{++}){}^{g_R}=V^{++}$,
$\Omega{}^{g_R} =\Omega g_R$. Because action (\ref{class}) of the
${\cal N}=2$ supersymmetric massless Yang-Mills field theory is
gauge invariant, the substitution $V^{++}\rightarrow
{}^{\Omega}V^{++}$ does not change the structure of action
(\ref{class}).

It is interesting to present another, explicitly gauge-invariant
but nonlocal form of the mass term expressed in terms of the
superfield strength $W$. We consider the action
\begin{equation}\label{class2}
S[V^{++},\omega]=-\frac{1}{2g^2}\mbox{tr}\int d^4\theta
W^2-\frac{m^2}{2g^2}\mbox{tr}\int d\zeta^{(-4)}du\,{\cal
V}^{++}{\cal V}^{++},
\end{equation}
where
\begin{equation}\label{cal_V}
{\cal V}^{++}=V^{++}-L^{++},
\end{equation}
$$
L^{++}=i(D^{++}\Omega) \Omega^{-1}=\int_0^1 d\tau e^{-i\tau\omega}
D^{++}\omega \, e^{i\tau\omega} =\int_0^1 d\tau D^{++}\omega_a
R_{ab}(\tau\omega)T_b$$ (we define the isotopic matrix
$R_{ab}(\omega)$ below; see (\ref{R})), and write the equation of
motion that follows from this action:
\begin{equation}\label{eguatmot}
\frac{1}{4}(D^+)^2W+m^2{\cal V}^{++}=0.
\end{equation}
Because $W$ is independent of harmonics, the consistency condition
for this equation is the equation of motion of the
$\omega$-multiplet:
\begin{equation}\label{constr}
D^{++}{\cal V}^{++}=0.
\end{equation}

As is clear from definition (\ref{cal_V}), the component content
of the gauge-covariant (${}^g{\cal V}^{++}=g{\cal V}^{++}g^{-1}$)
potential ${\cal V}^{++}$ is determined by complicated
nonpolynomial combinations composed from physical components of
the vector multiplet in the Wess-Zumino gauge and from the
components of the $\omega$-multiplet. But because of constraint
(\ref{constr}), we can parameterize the component decomposition on
the mass shell:
\begin{equation}\label{componentV}{\cal V}^{++}(\zeta, u)=f^{++}(x_A)+(\theta^+)^2\bar\varphi(x_A)
+(\bar\theta^+)^2\varphi(x_A)+2i(\theta^+\sigma^m\bar\theta^+)(A_m(x_A)+\partial_m
f^{+-}(x_A))
\end{equation}$$+2[\theta^{+\alpha}\psi^i_\alpha-
\bar\theta^+_{\dot\alpha}\bar\psi^{\dot\alpha
i}]u^+_i+2i[(\bar\theta^+)^2\theta^{+\alpha}\partial_{\alpha\dot\alpha}\bar\psi^{\dot\alpha
i}+(\theta^+)^2\bar\theta^+_{\dot\alpha}\partial^{\dot\alpha\alpha}\psi^i_\alpha]u^-_i-(\theta^+)^2(\bar\theta^+)^2\Box
f^{--}(x_A).
$$
The mass term for the vector component of the superfield ${\cal
V}^{++}$ then has a form similar to the form of the mass term in
the Stueckelberg formalism for the nonsupersymmetric massive
Yang-Mills field theory \cite{yukaf}:
$$S_m =-\frac{m^2}{2g^2} \mbox{tr}\int d^4x du (A_m - L_m)^2,$$
where the Cartan form $L_m$ on the group is determined in terms of
the isoscalar and isotriplet physical components $\omega,$ and
$\omega^{(ij)}$ of the $\omega$-supermultiplet. Deriving an
explicit component expression for $S_m$ is technically difficult
because we must integrate over harmonics in each term of the
infinite series for the Cartan form
$L_m=e^{-if^{ij}u^+_iu^-_j}\partial_m e^{if^{ij}u^+_iu^-_j}$ ,
where $f^{ij}=\omega\varepsilon^{ij}+\omega^{ij}.$ But the analogy
with the corresponding nonsupersymmetric theory is nevertheless
transparent.

We can treat Eq. (\ref{eguatmot}) as a constraint imposed on the
$\omega$-multiplet, which can also be resolved perturbatively in
the non-Abelian theory, but whose solution has an especially
simple form in the Abelian case:
\begin{equation}\label{solution}
\omega(\zeta_1, u_1)=\int
d\zeta^{(-4)}_2du_2G^{(0,0)}_0(1|2)D^{++}_2V^{++}(2),
\end{equation}
where
\begin{equation}\label{GreenOmega}
G^{(0,0)}_0(1|2)=-\frac{1}{\Box}(D^+_1)^4(D^+_2)^4\delta^{12}(1|2)\frac{u^-_1u^-_2}{(u^+_1u^+_2)^3}
\end{equation}
is the Green's function of the omega-multiplet \cite{gikos},
\cite{gios}. Acting with the operator $D^{++}$ on the both sides
of (\ref{solution}), we obtain : $$D^{++}_1\omega(\zeta_1,
u_1)=-\int d\zeta^{(-4)}_2du_2
\frac{1}{\Box}(D^+_1)^4(D^+_2)^4\delta^{12}(1|2)V^{++}(2)\{\frac{1}{(u^+_1u^+_2)^2}$$$$+\frac12(D_2^{--})^2\delta^{(2,-2)}(u_2,u_1)\}
=\int
d\zeta^{(-4)}_2du_2\{\Pi_T^{(2,2)}(1|2)+\delta^{(2,2)}_A(1|2)\}V^{++}(2),
$$
where $\Pi_T$ is an analytic distribution with the properties of
the projection operator \cite{gios}. Now excluding the gauge
degrees of freedom of $\omega$ from (\ref{class2}), after a chain
of transformations for the mass term, we obtain:
\begin{equation}\label{nonloc}
S_m=-\frac{m^2}{2g^2}\int
d\zeta_1^{(-4)}du_1d\zeta_2^{(-4)}du_2V^{++}(1)\Pi_T^{(2,2)}(1|2)V^{++}(2)=\frac{m^2}{2g^2}\int
d^8 z_c  W\frac{1}{\Box}W.
\end{equation}
As a result, the gauge-invariant form of the mass term can be
completely formulated in terms of the field strength  $W$. In a
non-Abelian case, we can obviously also find the
$\omega$-multiplet perturbatively as a polynomial in powers of
$V^{++}$. This expansion is definitely a nonlocal expression, but
we can localize it by introducing the proper tensor
multiplets,\footnote{We write the action for the massive tensor
multiplet in the harmonic superspace \cite{kuz_mass} in the form
$$S=\frac12\int d\zeta^{-4}(G^{++})^2+\frac12 m\{\int d^8z W\psi
+c.c.\}+\frac12\int d^8z W^2,$$ where $G^{++}(z,u)$ is the real
analytic superfield satisfying the equation $D^{++}G^{++}=0$. This
equation can be resolved in terms of the harmonic-independent
unconstrained chiral superfield $\psi(z)$ and its conjugate in the
form $G^{++}(z,u)=\frac18 (D^+)^2\psi(z)+\frac18
(\bar{D}^+)^2\bar\psi(z).$ This superfield remains invariant under
the gauge transformations $$\delta \psi=i\Lambda , \quad
\bar{D}^i_{\dot\alpha}\Lambda=0, \quad D^{\alpha i}D^j_{\alpha
}\Lambda = \bar{D}^i_{\dot\alpha}\bar{D}^{j\dot\alpha}\bar\Lambda.$$
We choose the gauge-fixing function in the form $F^{++}=\frac18
(D^+)^2\psi(z)-\frac18 (\bar{D}^+)^2\bar\psi(z).$ Integrating in the
generating functional over the prepotentials $\psi$ and $\bar\psi$,
$$Z=\int D\psi D\bar\psi (\mbox{Det}\Box)e^{\frac{i}{2}\int d
\zeta^{-4}\{ (G^{++})^2 +(F^{++})^2\}+\frac{i}{2} m\int d^8z
(W\psi +c.c.)}=e^{i\int d^8z W\frac{m^2}{\Box}W}$$ we obtain the
nonlocal mass term for vector potential (\ref{nonloc}).} and we
obtain an interesting sequence of classical dualities in the model
containing the Stueckelberg superfields $\omega$. An analogous
situation occurs when calculating the condensates $<A^2>$ in the
Yang-Mills theory \cite{zakharov}, where the nonlocal
gauge-invariant functional related to $A^2$ contains information
about the topological structure of the theory vacuum with a
nonvanishing mean of the operator (see \cite{sorella} and the
references therein for a detailed description of the infrared
dynamics of the ${\cal N}=0$ Yang-Mills theory).

Our further goal here is to determine the effective action and to
analyze the structure of one-loop divergences in the theory with
action (\ref{class2}).

\section{The background field formalism}
A gauge-invariant loop expansion of the effective action in
supersymmetric theories is given on the base of the superfield
background field method (see, e.g., \cite{idea} for ${\cal N}=1$
theories and \cite{backgr} for ${\cal N}=2$ theories). In the
background field formalism, we subsequently perform the
background-quantum splitting of all the fields, fix the gauge
degrees of freedom of quantum fields, and integrate only over
quantum fields in path integral. The contribution to the effective
action in a given loop order then comes from a finite number of
terms in expansion of action in the integrand in quantum fields.

We consider the theory of the fields $V^{++}$  and $\omega$ with
action (\ref{class2}). In the ${\cal N}=2$ sector of the vector
supermultiplet $V^{++}$, we split $V^{++}\rightarrow
V^{++}+gv^{++}$ and repeat all the steps as in the massless theory
\cite{backgr}. In the sector of the $\omega$-multiplet with a
nonlinear chiral Lagrangian, we must construct the expansion
following the perturbation theory in terms of
parameterization-independent invariant quantities \cite{volk}. The
main principle of the background-quantum splitting of fields
taking values in the group is a nonlinear rule for the group
addition of elements of the Lie group $\Omega(\omega)$ and
$\Omega(\chi)$ determined by the relation \cite{volk}:
\begin{equation}\label{omega}
\Omega(\omega \oplus \chi)= \Omega(\omega)\Omega(\frac{m}{g}\chi).
\end{equation}
Under such a rule $\Omega(\omega \oplus \chi)$ is an element in the
same space as $\Omega(\omega)$ and has the same group transformation
law as $\Omega(\omega)$.

We define the background-{quantum splitting of the superfield
$\omega$ into the background superfield $\omega$ and the quantum
superfield $\chi$ according to rule (\ref{omega}). It is easy to
demonstrate that the background $\omega$-fields transform as in
(\ref{transomeg}), while the quantum fields $\chi$ are merely
invariant. Under such a splitting of fields into background and
quantum fields, both the Lagrangian and all the terms of the
Taylor expansion in quantum fields are invariant under both the
local and global transformation groups. Consequently, all the
obtained counterterms are automatically invariant under classical
gauge and global transformations.

For the one-loop calculation, it is sufficient to expand the
Lagrangian up to terms of the second order in the quantum fields
$v^{++}$ and $\chi$:
\begin{equation}\label{S_2}
S^{(2)}=\frac12\int
d^{12}zdu_1du_2\frac{v_a^{++}(1)v_a^{++}(2)}{(u^+_1u^+_2)^2}
-\frac{1}{2}\int d\zeta^{(-4)}du \{m^2(v^{++}_a)^2
\end{equation}
$$
-2mD^{++}\chi_a R_{ab}v^{++}_b+D^{++}\chi_aD^{++}\chi_a
+f_{abc}D^{++}\chi_a\chi_b R_{cd}{\cal V}^{++}_d\},
$$
where the isotopic matrix $R_{ab}(\omega)$ is determined by the
equality $\Omega T_a\Omega^{-1}=R_{ab}T_b$. As in the ${\cal N}=0$
case \cite{yukaf}, it has the properties:
\begin{equation}\label{R}
R_{ae}R_{be}=\delta_{ab}, \quad
D^{++}R_{ab}=-R_{ae}f_{bec}L^{++}_c, \quad
f_{abc}=R_{ad}R_{be}R_{cg}f_{deg}.
\end{equation}
To (\ref{S_2}), we must also add the term fixing the gauge in the
sector of the quantum vector superfield, which can be conveniently
chosen in the background gauge-invariant form,
\begin{equation}
F^{(+4)}={\cal D}^{++}v^{++},
\end{equation}
and the action of the Faddeev-Popov and Nielsen-Kallosh ghosts.
Here, we follow the approach developed in \cite{backgr}. Following
the Faddeev-Popov procedure, to fix the gauge in the functional
integral $Z=N\int {\cal D}v^{++}e^{iS}$, we must insert unity in
the form $1=\Delta_{FP}\delta(F^{(+4)}-f^{(+4)})$, where the
Faddeev-Popov determinant is $\Delta_{FP}[v^{++},
V^{++}]=\mbox{Det}({\cal D}^{++}({\cal D}^{++}+iv^{++}))$.
Further, we must insert unity in the form
$$
1=\Delta_{NK}\int {\cal
D}f^{(+4)}\exp\{\frac{i}{2\alpha}\mbox{tr}\int d^{12}zdu_1du_2
f_\tau^{(+4)}\frac{(u^-_1u^-_2)}{(u^+_1u^+_2)^3}f_\tau^{(+4)}\},
$$
into the functional integral, where $\alpha$ is the gauge
parameter, which for convenience we set equal to $\alpha=-1$ in
what follows, $\Delta_{NK}[V^{++}]$ is the Nielsen-Kallosh
determinant, and $f_\tau^{(+4)}=e^{-ib}f^{(+4)}e^{ib}$ is a gauge-
invariant function taking values in the Lie algebra of the gauge
group. We note that the Nielsen-Kallosh determinant depends on the
background superfield, which indicates the presence of the third
Nielsen-Kallosh ghost. The details of calculation of this
determinant $\Delta_{NK}[V^{++}]=\mbox{Det}^{-1/2}({\cal
D}^{++})^2\mbox{Det}^{1/2}\stackrel{\frown}{\Box}_{(4,0)}$ are
given in \cite{backgr}. Here $\stackrel{\frown}{\Box}$ is the
covariant-analytic D'Alembertian transforming analytic superfields
again into analytic superfields \cite{backgr}:
\begin{equation}\label{smile}
\stackrel{\frown}{\Box}=-\frac12({\cal D}^+)^4({\cal D}^{--})^2
=\frac{1}{2}{\cal D}^{\alpha\dot\alpha}{\cal D}_{\alpha\dot\alpha}
+\frac{i}{2}({\cal D}^{+\alpha} W){\cal D}^-_\alpha
+\frac{i}{2}(\bar{\cal D}^{+}_{\dot\alpha}\bar{W})\bar{\cal
D}^{-\dot\alpha}
\end{equation}
$$
+\frac{1}{2}\{{W}, \bar{W}\}-\frac{i}{4}(\bar{\cal D}^+\bar{\cal
D}^+\bar{W}){\cal D}^{--} +\frac{i}{8}[{\cal D}^+,{\cal D}^-]{W}~.
$$

The final result for the Lagrangian determining one-loop quantum
corrections to the effective action in the vector multiplet sector
is
\begin{equation}
S_2 +S_{GF}=-\frac{1}{2}\mbox{tr}\int d\zeta^{(-4)}du
v^{++}(\stackrel{\frown}{\Box}+m^2)v^{++}.
\end{equation}
The ghost action is:
\begin{equation}\label{FPNK}
S_{ghost}=\mbox{tr}\int d\zeta^{(-4)}du b ({\cal
D}^{++})^2c+\frac{1}{2}\int d\zeta^{(-4)}du\phi ({\cal
D}^{++})^2\phi + \mbox{tr}\int d\zeta^{(-4)}du \rho^{(+4)}
\stackrel{\frown}{\Box}_{4,0}\sigma.
\end{equation}
where $b$ and $c$ are the anticommuting superfield Faddeev-Popov
ghosts, $\rho^{(+4)}$ and $\sigma$ are the anticommuting
Nielsen-Kallosh ghosts, and $\phi$ are additional commuting ghosts
taking values in the Lie algebra of the gauge group.

It is convenient to write the superfield action for the quantum
field $\chi$ and for its interaction with the quantum field
$v^{++}$ in the matrix form:
\begin{equation}\label{matrix}
S^{(2)}_{SYM}+S^{(2)}_m=-\frac{1}{2}\int d\zeta^{(-4)}(v^{++}_a,
\chi_a)\pmatrix{\stackrel{\frown}{\Box}_{ab}+m^2&-mR_{ba}D^{++}\cr
mD^{++}R_{ab}&-\nabla^{++}D^{++}}\pmatrix{v^{++}_b\cr \chi_b},
\end{equation}
where $D^{++}$ is the standard harmonic derivative \cite{gios} and
\begin{equation}\label{long} \nabla^{++} \chi_a=D^{++}\chi_a
+f_{abc}\chi_b R_{cd}{\cal V}^{++}_d
\end{equation}
is the long derivative in the $\lambda$-basis in which $\Omega {\cal
V}^{++}_aT_a \Omega^{-1}$ plays the role of the (gauge-invariant)
connection. Because this field is analytic, we have the standard
constraints $$[\nabla^{++}, \nabla^+_{\alpha,\; \dot\alpha}]=0.$$
For uniformity here and hereafter, we use the notation
$\nabla^+_{\alpha,\; \dot\alpha} = D^+_{\alpha,\; \dot\alpha}$.
Other commutation relations, for instance,
\begin{equation}\label{algebra}
[\nabla^{++}, \nabla^{--}]=D^0, \quad [\nabla^{\mp\mp},
\nabla^{\pm}_{\alpha, \,\dot\alpha}]=\nabla^{\mp}_{\alpha,
\,\dot\alpha},
\end{equation}
$$
\{\bar\nabla^+_{\dot\alpha},\,
\nabla^-_\alpha\}=-\{\nabla^+_\alpha,\,
\bar\nabla^-_{\dot\alpha}\}=2i\nabla_{\alpha\dot\alpha},
$$
$$
\{\nabla^+_\alpha,\,
\nabla^-_\beta\}=-2i\epsilon_{\alpha\beta}\bar{\cal W}, \quad
\{\bar\nabla^+_{\dot\alpha},\,
\bar\nabla^-_{\dot\beta}\}=2i\epsilon_{\dot\alpha\dot\beta}{\cal
W},
$$
exactly replicate the commutation relations for the covariant
derivatives ${\cal D}^{\pm}_{\alpha,\dot\alpha}$ and ${\cal
D}^{\pm\pm,0}$ \cite{backgr} and determine $\nabla^{-}_{\alpha,
\dot\alpha}$ and the chiral superfield of the harmonically
independent strength ${\cal W}[{\cal
V^{++}}]=-\frac14(\nabla^+)^2{\cal V^{--}}$ and $\bar{\cal
W}[{\cal V^{++}}]=-\frac14(\bar\nabla^+)^2{\cal V^{--}}$ for the
gauge invariant potential $\Omega{\cal V}^{++}\Omega^{-1}$.

\section{One-loop divergences}
We now analyze one-loop divergences in the theory under
consideration. The effective action is the sum of action
(\ref{FPNK}) of quantum superfields of ghosts and action
(\ref{matrix}) of the quantum superfields $v^{++}$ and $\chi$:
\begin{equation}
\Gamma^{(1)}[V^{++}, \omega] = \Gamma^{(1)}_{1}[V^{++}] +
\Gamma^{(1)}_{2}[V^{++}, {\cal V}^{++}].
\end{equation}
where $\Gamma^{(1)}_{1}[V^{++}]$ is the ghost contribution to the
effective action and $\Gamma^{(1)}_{2}[V^{++}, {\cal V}^{++}]$ is
the contribution from the superfields $v^{++}$ and $\chi$. Here,
${\cal V}^{++}$ is given by expression (\ref{cal_V}). We note that
the whole dependence of the effective action on the Stueckelberg
superfield $\omega$ is contained in ${\cal V}^{++}$ (\ref{cal_V}).
Actions (\ref{FPNK}) and (\ref{matrix}) completely determine the
structure of a perturbative expansion needed for calculating the
one-loop effective action of the massive ${\cal N}=2$ supersymmetric
Yang-Mills field theory in an explicitly supersymmetric and
gauge-invariant form. Further, we are interested only in the
structure of divergences of the considered theory. For this, we use
the dimensional regularization (see \cite{idea}) about using
dimensional regularization in superfield theories) and the minimal
subtraction scheme.

The ghost contribution to the one-loop effective action depends only
on the potential $V^{++}$ and coincides completely with the
corresponding contribution in the standard massless ${\cal N}=2$
supersymmetric Yang-Mills field theory \cite{backgr}:
\begin{equation}
i\Gamma^{(1)}_{1}[V^{++}]=\mbox{Tr} \ln({\cal
D}^{++})^2-\frac{1}{2}\mbox{Tr} \ln({\cal D}^{++})^2
+\frac{1}{2}\mbox{Tr}\ln\stackrel{\frown}{\Box}_{(4,0)}.
\end{equation}
We can therefore directly use the results in \cite{backgr},
\cite{effect} to calculate the divergent part of the effective
action:
\begin{equation}\label{gh contr}
\Gamma^{(1)}_{1,div}[V^{++}] =
-\frac{C_2}{32\pi^2\varepsilon}\mbox{tr}\int d^8z {W}^2,
\end{equation}
where $C_2$ is the quadratic Casimir operator of the gauge group
and $\varepsilon$ is the dimensional regularization parameter.

New contributions to divergences correspond to one-loop
corrections to the effective action related to the quantum fields
$v^{++}$ and $\chi$ running along the loop and to their mixing. To
calculate the functional determinant of the matrix operator in
action (\ref{matrix}), it is convenient reduce the matrix to the
diagonal form and write it as:
\begin{equation}\label{matrix2}
\pmatrix{1&mRD^{++}\frac{1}{\nabla^{++}D^{++}}\cr0&1}\pmatrix{\stackrel{\frown}{\Box}+m^2
-m^2RD^{++}\frac{1}{\nabla^{++}D^{++}}D^{++}R &0\cr
0&-\nabla^{++}D^{++}}
\end{equation}
$$
\times \pmatrix{1&0\cr -\frac{1}{\nabla^{++}D^{++}}mD^{++}R &1}.
$$
All the contributions to the effective action are then ensured by
the diagonal elements of the matrix:
\begin{equation}\label{diag}
\pmatrix{\stackrel{\frown}{\Box}+m^2\Pi^T&0\cr0&-\nabla^{++}D^{++}}.
\end{equation}
where we use the notation $\Pi^T$ for the covariant-analytic
distribution \cite{backgr} with the projection operator
properties. It is well known that the operator $
\stackrel{\frown}{\Box}+m^2 \Pi^T$ and also $
\stackrel{\frown}{\Box}_{(4,0)}$ do not contribute to the
holomorphic part of the effective action \cite{backgr},
\cite{effect}. All possible contributions to the one-loop
counterterm are then due to the known ghost contribution (\ref{gh
contr}) and due to the contribution
\begin{equation}\label{chi_contr}
\Gamma^{(1)}_{2}[0, {\cal V}^{++}]= \Gamma^{(1)}[{\cal
V}^{++}]=\frac{i}{2}\mbox{Tr}\ln(\nabla^{++}D^{++})
\end{equation}
of the quantum superfields $\chi$ coming from the lower-right
block of matrix (\ref{diag}). To use the known calculation tools
\cite{backgr}, \cite{effect} for $\Gamma^{(1)}[{\cal V}^{++}]$, we
reduce the differential operator in (\ref{chi_contr}) to the form
\begin{equation}
(\nabla^{++})^2+U^{(+4)},
\end{equation}
where
\begin{equation}
U^{(+4)}_{ab}=\frac12f_{abc}D^{++}{\cal V}^{++}_c+\frac14
f_{ace}f_{bde}{\cal V}^{++}_c{\cal V}^{++}_d.
\end{equation}
Obviously, the exact Green's function for the operator
$(\nabla^{++})^2$ coincides with (\ref{GreenOmega}) after the
replacement ${\cal D}^{++}\rightarrow \nabla^{++}$ and
$W\rightarrow{\cal W}$. The Green's function for the
omega-multiplet in the external field $U^{(+4)}$ is determined by
the equation
\begin{equation}
[(\nabla^{++}_1)^2+U_1^{(+4)}]G_U^{(0,0)}(1|2)=\delta_A^{(4,0)}(1|2).
\end{equation}
We define the analytic superfield kernel by the law
\begin{equation}
Q^{(4,0)}(1|2)=\delta_A^{(4,0)}(1|2) +U_1^{(+4)}G^{(0,0)}(1|2)
\end{equation}
where the Green's function $G^{(0,0)}(1|2)$ in the external field
${\cal V}^{++}$ satisfies the equation
$$(\nabla^{++}_1)^2G^{(0,0)}(1|2)=\delta^{(4,0)}(1|2).$$ This kernel
contains all the external field effects. The effective action is
then determined as
\begin{equation}\label{gammaV}
\Gamma[{\cal V}^{++}]=\frac{i}{2}\mbox{Tr}\ln
((\nabla^{++})^2+U^{(+4)})=\frac{i}{2}\mbox{Tr}\ln
(\nabla^{++})^2+\frac{i}{2}\mbox{Tr}\ln Q^{(4,0)}.
\end{equation}
The first term in the right-hand side of this equation exactly
coincides with the known one-loop contribution of the
hypermultiplet in an external field \cite{effect}. This easily
follows from comparing this contribution with expression (\ref{gh
contr}). The only difference is that we have the opposite sign and
replace the superfield strength $W[V^{++}]$  with ${\cal W}[{\cal
V^{++}}]=-\frac14(\bar\nabla^+)^2{\cal V}^{--}$. We can therefore
write the contribution to the divergent part of the effective
action additional to (26) without calculations:
\begin{equation}\label{calV_contr}
{}_{1}\Gamma_{div}^{(1)}[{\cal V}^{++}] =
\frac{C_2}{32\pi^2\varepsilon}\mbox{tr}\int d^8z {\cal W}^2.
\end{equation}

On the diagram level, the expansion of the second term in the
right-hand side of equality (\ref{gammaV}) in a power series in
interactions of fields inside the loop with the external
insertions $U^{(+4)}$ and with the propagator in the external
field ${\cal V }^{++}$ is
\begin{equation}\label{sum}
\Gamma[U^{(+4)}]=\sum_{n=1}^\infty\Gamma_n[U^{(+4)}],
\end{equation}
where the nth term of this series is described by the supergraph
with n external lines $U^{(+4)}$. Functional (\ref{gammaV})
therefore contains the complete information about one-loop
contributions with an arbitrary number of external lines ${\cal
V}^{++}$.

The first term in (\ref{chi_contr}) in the expansion of
$\Gamma[{\cal V}^{++}]$ in a power series in $U^{(+4)}$ vanishes
because it contains the harmonic product, which becomes zero in
the limit of coinciding arguments, $u^-_1u^-_2|_{u_1=u_2}=0$. The
effective action in the second order is
$$
{}_{2}\Gamma^{(1)}[{\cal V}^{++}]=-\frac{i}{4}\mbox{tr}\int
d\zeta^{(-4)}_1 d\zeta^{(-4)}_2du_1du_2
U^{(+4)}(1)U^{(+4)}(2)\frac{1}{\stackrel{\frown}{\Box}_1}(\nabla^+_1)^4(\nabla^+_2)^4
\delta^{12}(1|2)\frac{u^-_1u^-_2}{(u^+_1u^+_2)^3}
$$
$$
\times
\frac{1}{\stackrel{\frown}{\Box}_2}(\nabla^+_2)^4(\nabla^+_1)^4
\delta^{12}(2|1)\frac{u^-_2u^-_1}{(u^+_2u^+_1)^3}.
$$
Reconstructing the total Grassmann integration measure, we remove
one of the delta functions and use the identity
$(D^+_1)^4(D^+_1)^4\delta^8(\theta-\theta')|_{\theta=\theta'}=(u^+_1u^+_2)^4$.
Further, after standard transformations, the divergent
contribution becomes
\begin{equation}\label{gammaU}
{}_{2}\Gamma^{(1)}_{div}[{\cal
V}^{++}]=\frac{1}{(8\pi)^2\varepsilon}\mbox{tr}\int
d^{12}zdu_1du_2
U^{(+4)}(z,u_1)U^{(+4)}(z,u_2)\frac{(u^-_1u^-_2)^2}{(u^+_1u^+_2)^2}.
\end{equation}
Subsequent terms in the expansion of (\ref{chi_contr}) give finite
contributions to the effective action.

Relation (\ref{gammaU}) is the main result in this paper. It is a
new superfield counterterm in the ${\cal N}=2$ supersymmetric
massive Yang-Mills field theory in the Stueckelberg formalism
depending on the background superfield $\omega$. Obviously, this
functional does not contain on-shell harmonic singularities. This
follows because the nonzero contribution to the integral over odd
variables must contain the maximum power of the Grassmann
coordinates. But
$(\theta^+_1)^2(\bar\theta^+_1)^2(\theta^+_2)^2(\bar\theta^+_2)^2=(u^+_1u^+_2)^4(\theta)^4(\bar\theta)^4.$.
Among many terms arising in its component form, functional
(\ref{gammaU}) contains nonstandard contact four-vector interactions
and terms necessary for their supersymmetrization. For example, for
the gauge group $SU(2)$, these interactions are ai
$a^i_ma^i_na^j_ma^j_n +(a^i_ma^i_m)^2$, where $a_m=A_m-L_m$ is the
vector component of the $SU(2)$-superfield ${\cal V}^{++}$ given by
(\ref{cal_V}). For the gauge group $SU(3)$, the corresponding
interactions are $\frac56(a^a_ma^a_m)^2
+a^a_ma^a_na^b_ma^b_n-d_{abe}d_{ecd}a^a_ma^b_na^c_ma^d_n$. Precisely
this counterterm arises as an obstruction to the renormalizability
of the standard nonsupersymmetric massive Yang-Mills field theory
\cite{shiza}, \cite{yukaf}. Such deviations from the standard model
result in interesting phenomenological consequences, for instance,
for the processes of the creation of $W^+W^-$ and $WZ$ vector bosons
(see \cite{cvetic} and the references therein). We note that in
contrast to the nonsupersymmetric case \cite{yukaf}, the mass term
in the ${\cal N}=2$ supersymmetric massive Yang-Mills field theory
is not renormalized.

All counterterms (\ref{gh contr}), (\ref{calV_contr}), and
(\ref{gammaU})\footnote{In the minimal subtraction scheme,
counterterms differ only in signs from divergences (\ref{gh contr}),
(\ref{calV_contr}), and (\ref{gammaU}).} are gauge-invariant,
however the counterterms (\ref{calV_contr}) and (\ref{gammaU}) do
not reproduce the form of the initial Lagrangian. Similar to the
massive ${\cal N}=0$ nonsupersymmetric Yang-Mills field theory, the
massive ${\cal N}=2$ supersymmetric Yang-Mills field theory can then
be regarded only as an effective low-energy theory. In other words,
action (\ref{class2}) is not the most general ${\cal N}=2$
supersymmetric functional compatible with the local left and global
right gauge symmetries of the theory, and for the theory to be
renormalized in the modern sense (see, e.g., \cite{Weinberg}) in the
next order of the derivative expansion of the effective action, we
must therefore include new vertices induced by functionals
(\ref{calV_contr}) and (\ref{gammaU}).

\section{The component structure of ${\cal N}=2$ superfield functional  (\ref{calV_contr})}

A separate interesting problem is to study features of the
component expansion of the ${\cal N}=2$ superfield strength ${\cal
W}$ constructed on the base of the gauge-invariant potential
$\Omega{\cal V}^{++}\Omega^{-1}$, which includes an additional
degree of freedom in the vector multiplet due to the Stueckelberg
superfield ${\omega}$. Obviously, this gauge-invariant superfield
differs from the covariant chiral superfield  strength constructed
using the potential ${\cal V}^{++}$ only by  the $\Omega$
operators, which disappears under the trace.

As a rule, the ${\cal N}=2$ superfield formalism is most useful for
description of interacting off-shell supermultiplets. Passing to
component fields nevertheless requires excluding an infinite number
of auxiliary fields, which is a rather difficult technical problem.
Finding the component structure of counterterm (\ref{gh contr}) is
easy; its component form coincides with that of the classical action
of the ${\cal N}=2$ supersymmetric Yang-Mills field theory
\cite{gikos}, \cite{gios}. But the component form of
(\ref{calV_contr}) and (\ref{gammaU}) needs a special investigation
because potential (\ref{cal_V}) transforms as ${}^g{\cal
V}^{++}=e^{i\lambda}{\cal V}^{++}e^{-i\lambda}$ in contrast to the
transformation law for $V^{++}$, this transformation law does not
contain the term with the derivative $D^{++}\lambda$. The superfield
${\cal V}^{++}$ therefore contains nonremovable longitudinal degrees
of freedom in the vector field sector. In particular, this makes
imposing the Wess-Zumino gauge impossible for the field ${\cal
V}^{++}$. The problem of finding the component form of superfield
functional (\ref{calV_contr}) therefore implies the component
decomposition of ${\cal W}$ without imposing a gauge condition on
${\cal V}^{++}$. A general solution of this problem is still missing
from the literature. Particular aspects of the component structure
of the massive vector supermultiplet without fixing a supergauge
were studied in \cite{volkmass} for the Abelian case and for the
non-Abelian case in the first order in the coupling constant
expansion.

In this section, we describe the procedure for finding the component
form of superfield functional (\ref{calV_contr}) in the bosonic
sector, in which the component content of the superfield ${\cal
V}^{++}$ necessary for writing (\ref{calV_contr}) and (\ref{gammaU})
in terms of physical fields actually coincides with the component
structure of the superfield $V^{++}$ in the Wess-Zumino gauge, but
each component is endowed with an infinite tower of interactions
with the longitudinal degrees of freedom, related to the component
fields of the $\omega$-multiplet.

A convenient way to find the component content of the strength
superfield for nonstandard theories \cite{ferrara} is based on
solving the harmonic zero-curvature equation \cite{gikos},
\cite{gios} for a nonanalytic potential ${\cal V}^{--}$:
\begin{equation}\label{zerocurve}
D^{++}{\cal V}^{--}-D^{--}{\cal V}^{++}+i[{\cal V}^{++}, {\cal
V}^{--}]=0.
\end{equation}
Because ${\cal V}^{++} = V^{++} - L^{++}$ and $V^{++}$ is subject to
the standard gauge transformations, we can impose the Wess-Zumino
gauge on $V^{++}$. As the result, gauge-invariant analytic potential
(\ref{cal_V}) takes on form (\ref{componentV}):
\begin{equation}\label{com_calV}
{\cal V}^{++}(\zeta)=f^{++}(x_A)+(\theta^+)^2\bar\varphi(x_A)
+(\bar\theta^+)^2\varphi(x_A)
+2i(\theta^+\sigma^m\bar\theta^+)(A_m(x_A)+\partial_m f^{+-}(x_A))
\end{equation}
$$-(\theta^+)^2(\bar\theta^+)^2\Box f^{--}(x_A) +
fermions,
$$
where $\zeta=(x^m_A, \theta^{+\alpha},\bar\theta^{+\dot\alpha},
u^{\pm}_i)$ and we preserve the notation for the component form of
the potential in the Wess-Zumino gauge. We can then remember that
each of the components of $V^{++}_{WZ}$ is endowed with an infinite
series in powers of the interaction with components of the
$\omega$-multiplet $ \omega(\zeta)=
\omega(x_A)+\omega^{(ij)}(x_A)u^+_iu^-_j $. Because the superfield
strength
\begin{equation}\label{cal_W}
{\cal W}=-\frac14(\bar{D}^+)^2{\cal V}^{--}
\end{equation}
is a harmonic-independent ${\cal N}=2$ chiral superfield, it is
convenient to seek a solution of (\ref{zerocurve}) in the chiral-
analytic coordinates $Z_c=(z_c,\bar\theta^{\pm\dot\alpha})$, where
$$z_c=(x^m_L,\theta^{\pm\alpha}), \quad x^m_L=x^m_A+2i\theta^-\sigma^m\bar\theta^+ .$$
In this basis, each of the components of potential (\ref{com_calV})
decomposes as:
$$
f(x_A)=f(x_L)-2i\theta^-\sigma^m\bar\theta^+\partial_mf(x_L)+(\theta^-)^2(\bar\theta^+)^2\Box
f(x_L).
$$
Following \cite{ferrara}, it is convenient to represent the
decomposition of ${\cal V}^{--}$ in these coordinates in the form
$$
{\cal V}^{--}(Z_c,u)=v^{--}(x_L, \theta^\pm,
u)+\bar\theta^+_{\dot\alpha}v^{(-3)\dot\alpha}+\bar\theta^-_{\dot\alpha}v^{-\dot\alpha}+(\bar\theta^-)^2{\cal
A}+(\bar\theta^+\bar\theta^-)\varphi^{--}+\bar\theta^{-\dot\alpha}\bar\theta^{+\dot\beta}\varphi^{--}_{\dot\alpha\dot\beta}
$$
$$
+(\bar\theta^+)^2v^{(-4)}+(\bar\theta^-)^2\bar\theta^+_{\dot\alpha}\tau^{-\dot\alpha}
+(\bar\theta^+)^2\bar\theta^-_{\dot\alpha}\tau^{(-3)\dot\alpha}+(\bar\theta^+)^2(\bar\theta^-)^2\tau^{--}.
$$
We note that strength (\ref{cal_W}) depends only on the fields $
{\cal A}$, $\tau^{-\dot\alpha}$, and $\tau^{--}$, buy not on all the
chiral superfields in this decomposition. However, as shown in
\cite{ferrara}, only the chiral superfield
$$ {\cal A}={\cal
A}_1 +(\theta^-)^2{\cal A}^{++}_4+(\theta^-\theta^+){\cal A}_5
+\theta^{-\alpha}\theta^{+\beta}{\cal
A}_{6\alpha\beta}+(\theta^+)^2{\cal
A}_7^{--}+(\theta^-)^2(\theta^+)^2{\cal A}_{10}+fermions
$$
actually determines the component structure of superfield functional
(\ref{calV_contr}), which has the form:
\begin{equation}\label{action}
S_{bos}=\frac14\mbox{tr}\int d^4x_Ldu\{2{\cal A}_1{\cal
A}_{10}+2{\cal A}^{++}_4{\cal A}^{--}_7-\frac12{\cal
A}^2_5-\frac14{\cal A}^2_6\}
\end{equation}
in the bosonic sector.

The equations that determine the components of the superfield ${\cal
A}$ are constructed as coefficients in expansion of
(\ref{zerocurve}) in $\theta^{\pm},\bar\theta^{\pm}$:
\begin{equation}\label{cal_A}
 \quad d^{++}{\cal A}_1=0,\quad  d^{++}{\cal
A}^{++}_4=0, \quad  d^{++}{\cal A}_5 +4{\cal A}^{++}_4=0, \quad
 d^{++}{\cal A}_{6\alpha\beta}=0
\end{equation}
$$
 \quad d^{++}{\cal A}^{--}_7 +2{\cal A}_5 +[\bar\phi, {\cal
A}_1]=0,  \quad  d^{++}{\cal A}_{10}+[\bar\phi, {\cal
A}^{++}_4]=0,
$$
where we introduce the notation
$$
d^{++}= \partial^{++}+i[f^{++}, ...].
$$
The general solution of this set of harmonic equations can be
written in terms of Green's functions for the operator
$\partial^{++}$ \cite{gios}. For example, the solution of the first
equation in chain (\ref{cal_A}) is:
\begin{equation}\label{A_1}
{\cal A}_1(u)=\varphi -i\int du_1
\frac{u^+u^-_1}{u^+u^+_1}[f^{++}(u_1), {\cal A}_1(u_1)]
\end{equation}
$$
=\sum_{n=0}^\infty (-i)^n\int du_1 ... du_n
\frac{u^+u^-_1}{u^+u^+_1} ...
\frac{u^+_{n-1}u^-_n}{u^+_{n-1}u^+_n}[ f^{++}(u_1),[ ...
,[f^{++}(u_n), \varphi]]]=e^{ib}\varphi,
$$
where  $\varphi$ is a particular solution of the homogeneous
equation and $e^{ib}$ is a nonanalytic superfield called the bridge
\cite{gikos}, \cite{gios} for the field
$f^{++}=-ie^{ib}\partial^{++}e^{-ib}$. This solution can be
constructed iteratively using the Taylor series in $f^{++}$ as in
\cite{gikos}.

Each component of $ {\cal A}$ is therefore determined by an infinite
series in powers of interactions of the standard components of the
superfield  $W$ with the field $f^{++}$:
\begin{equation}\label{A_all}
{\cal A}^{++}_4 =e^{ib}\Box f^{++}, \quad {\cal A}^{--}_7
=4e^{ib}\Box f^{--}, \quad {\cal A}_6^{\alpha\beta}= e^{ib}
F^{\alpha\beta}
\end{equation}
$$
{\cal A}_5=e^{ib}(-4\Box f^{+-}+\frac12[\varphi,\bar\varphi]),
\quad {\cal A}_{10}=e^{ib} (-\Box \bar\varphi
+\frac18[\bar\varphi,[\bar\varphi,\varphi]]).
$$
Here, the component fields $f^{ij}, \varphi, \bar\varphi$ and
$F^{\alpha\beta}$ are determined by decomposition (\ref{com_calV})
and the action of the matrix operator $e^{ib}$ is given by relation
(\ref{A_1}).

Our analysis therefore proves that a formal solution for the
components of superfield strength (\ref{cal_W}) exists in a
nonpolynomial form and that in addition to the standard action in
the Wess-Zumino gauge modified by interaction with the
$\omega$-multiplet components, action (\ref{action}) contains fourth
powers of the space-time derivatives of $\omega, \omega^{(ij)}$
components of the $\omega$-multiplet. Finding a more detailed
component form of expression (\ref{action}) is a very complicated
technical problem although in principle such a form can be found
using the above procedure. We also note that in the sector of the
vector multiplet components (i.e., when we switch off the dependence
on the omega-multiplet components), divergences (\ref{gh contr}) and
(\ref{calV_contr}) cancel.

\section{Conclusion}

We present the main results of the paper.

{\bf 1.} We considered the ${\cal N}=2$ supersymmetric massive
Yang-Mills field theory whose action depends on the ${\cal N}=2$
gauge superfield $V^{++}$ and on the hypermultiplet Stueckelberg
superfield $\omega$. We proposed the various dual-equivalent
formulations of this theory differing by form gauge-invariant mass
term in the superfield action.

{\bf 2.} We developed a background field method that allows
obtaining the loop expansion of the effective action in an
explicitly gauge-invariant and ${\cal N}=2$ supersymmetric form. We
demonstrated that the contribution of the Stueckelberg superfield
${\omega}$ to the effective action can be formulated in terms of the
superfield  ${\cal V}^{++}$ given by (\ref{cal_V}), which is a
special gauge-invariant combination of the background superfields
$V^{++}$ and ${\omega}$.

{\bf 3.} We studied the structure of one-loop divergences in the
theory under consideration. We obtained explicitly gauge-invariant
and ${\cal N}=2$ supersymmetric expressions for one-loop
divergences (\ref{gh contr}), (\ref{calV_contr}) and
(\ref{gammaU}). The expression (\ref{gammaU}) is a new
gauge-invariant and ${\cal N}=2$ supersymmetric functional
constructed from the superfields $V^{++}$ and ${\omega}$. The
appearance of this functional as an expression determining
one-loop divergences results in the (multiplicative)
nonrenormalizability of the theory. This functional can be treated
as an ${\cal N}=2$ supersymmetrization of the covariant
counterterm in the nonsupersymmetric Yang-Mills field theory
\cite{shiza}, \cite{yukaf}. But in contrast to the
nonsupersymmetric case, the mass term in the massive ${\cal N}=2$
supersymmetric Yang-Mills field theory is not renormalized. We
gaved the complete analysis of the ${\cal N}=2$ superfield
structure of the one-loop divergences in the considered theory.
The case when the interaction of the vector multiplet with the
Stueckelberg multiplet is switched off or, in other words, the
zero-mass case,  partly verifies the obtained results. In this
limit, we have a single divergence due to one-loop ghost
contributions (\ref{gh contr}), which determines the known value
of the beta-function of the pure ${\cal N}=2$ supersymmetric
Yang-Mills field theory \cite{effect}.

{\bf 4.} We considered the component structure of the counterterm of
form (\ref{calV_contr}) in the bosonic sector. Because the gauge
transformation for ${\cal V}^{++}$ given by (\ref{cal_V}) does not
contain the derivative of the gauge parameter, we cannot impose the
Wess-Zumino gauge on ${\cal V}^{++}$, which is standardly used when
passing from the superfield description of the vector multiplet to
its component description. We therefore encountered the problem of
finding the component form of superfield functional
(\ref{calV_contr}). A procedure for solving this problem in the
bosonic sector is proposed (see (\ref{action}) and (\ref{A_all})).

We briefly discuss the prospect for further studying the massive
${\cal N}=2$ supersymmetric Yang-Mills field theory. As is known,
the massless ${\cal N}=2$ supersymmetric Yang-Mills theory is finite
beyond the one-loop approximation (see, e.g., \cite{backgr}). The
problem of divergences of the considered massive theory in higher
loops remains open. Finding finite contributions to the one-loop
effective action and studying the effective action in the presence
of the interaction between the massive gauge ${\cal N}=2$ superfield
and the matter hypermultiplets is therefore interesting. In our
opinion, the problem of the quantum equivalence of the massive
${\cal N}=2$ supersymmetric Yang{Mills field theory in the
Stueckelberg formalism and the ${\cal N}=2$ supersymmetric
non-Abelian vector-tensor model, which (as was shown) are dual on
the classical level, is especially interesting.

\begin{center}

{\bf Acknowledgements}
\end{center}
One of the authors (N. G. P.) is grateful to I. Samsonov for the
useful discussion. This work was supported in part by the Russian
Foundation for Basic Research (Grant No. 06-02- 16346 and
08-02-00334-a), INTAS (Grant No. 05-7928), and the Programm for
Supporting Leading Scientific Schools (Grant No. NSh-2553.2008.2).

\end{document}